\documentclass[11pt]{article}
\usepackage{amsmath,amssymb,amsthm}
\usepackage{url}

\newcommand*{\fd}
[2]{\mathchoice{\frac{\delta#1}{\delta#2}}
  {\delta#1/\delta#2}{\delta#1/\delta#2}
  {\delta#1/\delta#2}}
\newcommand{\ddx}[1]{\partial_x^{#1}}
\newtheorem{theorem}{Theorem}

\newtheorem{corollary}[theorem]{Corollary}

\begin{document}

\title{Systems of conservation laws with third-order Hamiltonian structures}
\author{E.V. Ferapontov$^{1}$, M.V. Pavlov$^{2,3}$, R.F. Vitolo$^4$ \\
  % EndAName
 [3mm] $^{1}$Department of Mathematical Sciences,\\
  Loughborough University,\\
  Loughborough, Leics, UK\\
  \texttt{e.v.ferapontov@lboro.ac.uk}\\
  [3mm] $^{2}$Department of Mechanics and Mathematics,\\
  Novosibirsk State University,\\
  Pirogova Street 2, 630090 Novosibirsk, Russia\\
   % \texttt{mpavlov@itp.ac.ru}\\
    [3mm]$^{3}$Sector of Mathematical Physics,\\
  Lebedev Physical Institute of Russian Academy of Sciences,\\
  Leninskij Prospekt 53, 119991 Moscow, Russia\\
  \texttt{m.v.pavlov@lboro.ac.uk}\\
  [3mm] $^{4}$Department of Mathematics and Physics \textquotedblleft E. De
  Giorgi\textquotedblright ,\\
  University of Salento, Lecce, Italy\\
  and Sezione INFN di Lecce\\
  \texttt{raffaele.vitolo@unisalento.it} } \date{}
\maketitle

\begin{abstract}
  We investigate $n$-component systems of conservation laws that possess
  third-order Hamiltonian structures of differential-geometric type.
  % Examples include equations of associativity of two-dimensional topological
  % field theory (WDVV equations).  and various equations of Monge-Amp\`ere
  % type.
  The classification of such systems is reduced to the projective
  classification of linear congruences of lines in $\mathbb{P}^{n+2}$
  satisfying additional geometric constraints.  Algebraically, the problem can
  be reformulated as follows: for a vector space $W$ of dimension $n+2$,
  classify $n$-tuples of skew-symmetric 2-forms $A^{\alpha} \in \Lambda^2(W)$ such
  that
$$
\phi_{\beta \gamma}A^{\beta}\wedge A^{\gamma}=0,
$$
for some non-degenerate symmetric $\phi$.

\bigskip


\bigskip

\end{abstract}

\tableofcontents

\newpage

\newpage

\section{Introduction}

\subsection{Systems of conservation laws and line congruences in projective
  space}

Systems of conservation laws are $n$-component first-order PDEs of the form
\begin{equation} \label{eq:61} u^i_t = (V^i(\mathbf{u}))_x,
\end{equation}
$i=1, \dots, n$, where $V^i(\mathbf{u})$ is a (nonlinear) vector of fluxes. We
will assume that the characteristic speeds of system (\ref{eq:61}), that is,
the eigenvalues of the Jacobian matrix of the fluxes $V^i$, are real and
distinct (condition of strict hyperbolicity). Systems of conservation laws
appear in a wide range of applications in continuum mechanics and mathematical
physics, see e.g.  \cite{Jef, Lax, Sev, Roz, Ser, SV14}.
Following the geometric approach of \cite{AF1, AF2}, with system (\ref{eq:61})
we associate a congruence (that is, $n$-parameter family of lines),
\begin{equation}
  y^i=u^i y^{n+1}+V^iy^{n+2},
  \label{cong}
\end{equation}
in auxiliary projective space $\mathbb{P}^{n+1}$ with homogeneous coordinates
$(y^1:\cdots :y^{n+2})$.
% $(y^1:...:y^n:y^{n+1}:y^{n+2})$.  In the case $n=2$ we obtain a two-parameter
% family, or a congruence of lines in $P^3$. In the 19th century the theory of
% congruences was one of the most popular chapters of classical differential
% geometry
It was demonstrated in \cite{AF1, AF2} that various standard concepts of the
theory of conservation laws such as rarefaction curves, shock curves, linear
degeneracy, reciprocal transformations, etc, acquire a simple interpretation in
the language of the projective theory of congruences. In particular,
reciprocally related systems (\ref{eq:61}) correspond to projectively
equivalent congruences (\ref{cong}). Algebro-geometric aspects of this
correspondence were investigated in \cite{Mez1, Mez2, Mez3, MM}.

In this paper we utilise the above geometric correspondence for the
classification of systems (\ref{eq:61}) possessing third-order Hamiltonian
structures. We will show that congruences associated with Hamiltonian systems
are necessarily {\it linear}, that is, they are specified by $n$ linear
relations among the Pl\"ucker coordinates (in geometric language, codimension
$n$ linear sections of the Grassmannian $G(1, n+1)$). We recall that the lines of a linear
congruence in $\mathbb{P}^{n+1}$ can be characterised geometrically as
$n$-secants of the {\it focal variety} (jump locus), which is a codimension two
subvariety in $\mathbb{P}^{n+1}$ (possibly, reducible):
\begin{itemize}
\item For $n=2$ the focal variety of a linear congruence consists of 2 skew
  lines in $\mathbb{P}^3$.
\item For $n=3$ the focal variety of a generic linear congruence is a
  projection of the Veronese surface $V^2\subset \mathbb{P}^5$ into
  $\mathbb{P}^4$ \cite{Cas}.
\item For $n=4$ the focal variety is a Palatini threefold in $\mathbb{P}^5$
  \cite{Pal}, etc.
\end{itemize}
% Note that linearity of the congruence does not mean linearity of the system
In parametrisation (\ref{cong}), the Pl\"ucker coordinates are just
$u^i, V^i, u^iV^j-u^jV^i$.  Imposing $n$ linear relations among the Pl\"ucker coordinates
% \begin{equation}
%   a^{\alpha}_{ij}(u^iV^j-u^jV^i)+b^{\alpha}_iV^i+c^{\alpha}_{i}u^i+d^{\alpha}=0, \label{linear} \end{equation}
% $\alpha=1, \dots, n$,
we obtain a linear system for the fluxes $V^i$ which  implies
that $V^i$ are rational in $u$. Systems associated with linear congruences are
linearly degenerate, and satisfy the Temple property \cite{AF2}.
% The key examples come from the theory of equations of associativity of
% topological field theory (WDVV equations), see Section \ref{sec:ex} below.

\subsection{Third-order Hamiltonian operators}

Third-order Hamiltonian operators of differential-geometric type were
introduced by Dubrovin and Novikov in \cite{DN2}, and subsequently investigated
in \cite{GP91, Doyle, GP97, OM98, BP, fpv, fpv1}. They are defined by the
general formula
\begin{multline*}
  P^{ij}=g^{ij}\ddx{3}+b_{k}^{ij}u_{x}^{k}\ddx{2}
  +(c_{k}^{ij}u_{xx}^{k}+c_{km}^{ij}u_{x}^{k}u_{x}^{m})\ddx{}
  \\
  +d_{k}^{ij}u_{xxx}^{k}+d_{km}^{ij}u_{xx}^{k}u_{x}^{m}
  +d_{kmp}^{ij}u_{x}^{k}u_{x}^{m}u_{x}^{p},
\end{multline*}
where $u^i$, $i=1$, \dots, $n$, are the dependent variables, and the
coefficients $g^{ij}, \dots, d_{kmn}^{ij}$ are functions of $u^{i}$ only;
$\ddx{}$ stands for the total derivative with respect to $x$. The requirement
that the corresponding Poisson bracket,
\begin{displaymath}
  \{F, G\} = \int \fd{F}{u^i}P^{ij}\fd{G}{u^j} dx,
\end{displaymath}
is skew-symmetric and satisfies the Jacobi identities, imposes strong
constraints on the coefficients of $P$.
% , see \cite{Dorf}.
We restrict our considerations to the non-degenerate case, $\det g^{ij}\neq 0$;
in what follows we use $g^{ij}$ for raising and lowering indices. It was
demonstrated in \cite{GP91, Doyle} that there exists a coordinate system (flat
coordinates) in which Hamiltonian operator $P$ takes a simple factorised form
\cite{OM98},
\begin{equation}
  P^{ij}=\ddx{}\left(g^{ij}\ddx{}+c_{k}^{ij}u_{x}^{k}\right)\ddx{}.  \label{casimir}
\end{equation}
In what follows we will always work in the flat coordinates, and keep for them
the notation $u^i$; note that $u^i$ are nothing but the densities of Casimirs
of the corresponding Hamiltonian operator. Introducing
$c_{ijk}=g_{iq}g_{jp}c_{k}^{pq}$ one can show \cite{GP97} that the
skew-symmetry conditions and the Jacobi identities for operator (\ref{casimir}) are
equivalent to
% we have a nonlinear system which expresses the skew-symmetry and the Jacobi
% property of the Poisson brackets:
\begin{subequations}\label{eq3}
  \begin{gather}
    g_{mp,k}=-c_{mpk}-c_{pmk},\label{eq4}\\
    c_{mpk}=-c_{mkp},\label{eq5}
    \\
    c_{mpk}+c_{pkm}+c_{kmp}=0,\label{eq6}\\
    c_{mpk,l}=-g^{sq}c_{sml}c_{qpk}.  \label{eq7}
  \end{gather}
\end{subequations}
Equations (\ref{eq4})--(\ref{eq6}) imply \cite{fpv}
\begin{equation}
  c_{skm}=\frac{1}{3}(g_{sm,k}-g_{sk,m}).  \label{eqc} 
\end{equation}
The elimination of $c$ from equations (\ref{eq3}) gives a system for the metric
$g$,
\begin{subequations}\label{g}
  \begin{gather}
    g_{mk,p}+g_{kp,m}+g_{mp,k}=0, \label{g1}\\
    g_{m[k,p]l}=-\frac{1}{3}g^{sq} g_{s[l,m]}g_{q[k,p]}. \label{g2}
  \end{gather}
\end{subequations}
Equations (\ref{g1}) mean that $g$ is a Monge metric, and as such is an object
of projective differential geometry.  Building on the correspondence of Monge
metrics to quadratic complexes of lines in $\mathbb{P}^n$, in \cite{fpv, fpv1}
we proposed a classification of Hamiltonian operators (\ref{casimir}) for
$n\leq 4$.

In what follows we will also need the result of Balandin and Potemin \cite{BP}
according to which the general solution of system (\ref{g}) is given by the
formula
\begin{equation}
  g_{ij}=\phi _{\beta \gamma }\psi _{i}^{\beta }\psi _{j}^{\gamma }, \label{tri}%
\end{equation}
where $\phi _{\beta \gamma }$ is a non-degenerate constant symmetric matrix,
and
\begin{equation}
  \psi _{k}^{\gamma }=\psi _{km}^{\gamma }u^{m}+\omega _{k}^{\gamma }; \label{lin}%
\end{equation}
here $\psi _{km}^{\gamma }$ and $\omega _{k}^{\gamma }$ are constants such that
$\psi _{km}^{\gamma }=-\psi _{mk}^{\gamma }$. These constants have to satisfy
an additional set of quadratic relations,
\begin{equation}
  \phi _{\beta \gamma }(\psi _{is}^{\beta }\psi _{jk}^{\gamma }+\psi
  _{js}^{\beta }\psi _{ki}^{\gamma }+\psi _{ks}^{\beta }\psi _{ij}^{\gamma
  })=0, \label{ab}%
\end{equation}%
\begin{equation}
  \phi _{\beta \gamma }(\omega _{i}^{\beta }\psi _{jk}^{\gamma
  }+\omega _{j}^{\beta }\psi _{ki}^{\gamma }+\omega _{k}^{\beta }\psi
  _{ij}^{\gamma })=0, \label{zac}%
\end{equation}
whose algebraic meaning was clarified in \cite{fpv1}. An important invariant of
Hamiltonian operator (\ref{casimir}) is its singular variety, $\det g_{ij}=0$,
which, due to (\ref{tri}), is a double hypersurface of degree $n-1$:
$$
\det g=\det \phi (\det \psi)^2;
$$
here the degree of $\det \psi$ equals $n-1$ \cite{fpv1}.

\subsection{Hamiltonian systems of conservation laws}

In this paper we are interested in  {\it Hamiltonian}
systems of conservation laws, 
namely systems (\ref{eq:61})
% $$ u_{t}^{i}=V^{i}(\mathbf{u})_{x}, $$
 possessing third-order Hamiltonian structures (\ref{casimir}):
$$
{\bf u}_t=(V({\bf u}))_x=P\frac{\delta H}{\delta {\bf u}},
$$
for some (nonlocal) Hamiltonian functionals $H$.
% $$ P=\ddx{}\left(g^{ij}\ddx{}+c_{k}^{ij}u_{x}^{k}\right)\ddx{}.  $$
Examples of such systems include Monge-Amp\`ere equations, as well as various
versions of WDVV equations, see \cite{Dub2b} for their geometric treatment
based on the theory of Frobenius manifolds. Our main results in this direction
can be summarised as follows.
% Note that the corresponding Hamiltonian density will automatically be
% nonlocal.  In this case condition~\eqref{eq:71} can be computed explicitly,
% leading to the following result.

\begin{theorem} \label{ThV} The necessary and sufficient conditions for a
  conservative system (\ref{eq:61}) to possess third-order Hamiltonian operator
  (\ref{casimir}) are the following:
  \begin{subequations}
    \label{V}
    \begin{align}\label{e1}
      & g_{im}V^{m}_{j}=g_{jm}V^m_{i},\\
      \label{e3}
      & c_{mkl}V^m_{i}+c_{mik}V^m_{l}+c_{mli}V^m_{k}=0,\\
      \label{e2}
      &V^k_{ij}=g^{ks}c_{smj}V^m_{i}+g^{ks}c_{smi}V^{m}_{j},
    \end{align}
  \end{subequations}
  here lower indices of $V^m$ denote partial derivatives:
  $V^m_i=\partial V^m/\partial u^i$, etc.
\end{theorem}
In Theorem \ref{Hamilt} of Section \ref{sec:Ham} we present explicit formulae
for the corresponding Casimirs, Momentum and Hamiltonian. The proof of Theorem
\ref{ThV} can be found in Section \ref{sec:ThV}.  

Conditions (\ref{V}) are  analogous to Tsarev's conditions in the theory of first-order
homogeneous Hamiltonian operators \cite{st}.
System (\ref{V}) possesses a
number of important properties, in particular, in Section \ref{sec:inv} we
establish the following result:

\begin{theorem} \label{Thinv} System \eqref{V} is in involution. Its general
  solution depends on $\leq \frac{n(n+3)}{2}$ arbitrary constants.
\end{theorem}

It is quite remarkable that system (\ref{V}), which is a linear involutive
system with non-constant coefficients, can be integrated in closed form
(Section \ref{sec:sol}).
% The key observation is that the quantities $\psi^{\gamma}_{k}V^{k}$, where
% $\psi^{\gamma}_{k}$ is the non-degenerate matrix from (\ref{tri}), must be
% linear in $u$.  system (\ref{V}) takes the
% form \begin{subequations} \begin{align} \nonumber &
%     \phi_{\beta\gamma}[\psi_{ik}^{\beta}W^{\gamma}+\psi_{k}^{\beta}W_{i}^{\gamma
%   }-\psi_{i}^{\beta}W_{k}^{\gamma}]=0,\\ \nonumber &
%     \phi_{\beta\gamma}[\psi_{ij}^{\beta}W_{k}^{\gamma}+\psi_{jk}^{\beta}% W_{i}^{\gamma}+\psi_{ki}^{\beta}W_{j}^{\gamma}]=0,\\ \nonumber &W_{ij}^{\gamma}=0 \end{align} \end{subequations} The last equation implies $W^{\gamma}$ are linear functions, $W^{\gamma}=w_{m}^{\gamma}u^{m}+\xi^{\gamma}$,
This leads to the following result (Section \ref{sec:sol}):

\begin{theorem}\label{ThHam} For Hamiltonian system (\ref{eq:61}) the following
  conditions hold:

  \begin{itemize}
  \item The associated congruence (\ref{cong}) is linear.
  \item System (\ref{eq:61}) is linearly degenerate and belongs to the Temple
    class.
  \item The fluxes $V^i$ are rational functions of the form
$$
V^i=\frac{Q^i}{\det \psi},
$$
where $\det \psi$ is a polynomial of degree $n-1$ defining the singular
variety, and $Q^i$ are polynomials of degree $n$.

 \end{itemize}
 Note that for $n\geq 4$ systems of conservation laws possessing third-order
 Hamiltonian structures are neither diagonalisable nor integrable in general.
\end{theorem}

Based on the classification of linear congruences in $\mathbb{P}^3$ and
$\mathbb{P}^4$ dating back to the classical work of Castelnuovo \cite{Cas},
this leads to a complete description of Hamiltonian systems (\ref{eq:61}) for
$n=2, 3$, see Section \ref{sec:class}.
% (Section \ref{sec:clas}):

\subsection{Examples}
\label{sec:ex}

Here we list examples of conservative systems (\ref{eq:61}) with third-order
Hamiltonian structures (\ref{casimir}) that will feature in the classification
results below.  In order to
simplify the expressions for the Hamiltonian densities  we introduce  potential coordinates  $b^i$ via $u^i=b^i_x$. In these coordinates  system~\eqref{eq:61} will  no longer be quasilinear, and  third-order Hamiltonian operator (\ref{casimir}) takes a
first-order form, see \eqref{eq:24}, \eqref{eq:500}.

\medskip

\noindent{\bf Example 1.} A linear $n$-component system of conservation laws,
$$
u^i_t=(a^i_ju^j)_x,
$$
$a^i_j=\text{const}$, possesses third-order Hamiltonian formulation
$$
u^i_t=\eta^{ij}\ddx{3}\frac{\delta H}{\delta u^j},
$$
$\eta^{ij}=\text{const}$, with the nonlocal Hamiltonian
$${ H}=-\frac{1}{2}\int \eta_{jp}a^p_kb^jb^k \ dx.
$$
In this case, conditions (\ref{V}) reduce to $\eta_{jp}a^p_k=\eta_{kp}a^p_j$,
which means that the operator $a$ is symmetric with respect to the metric
$\eta$.

The associated congruence (\ref{cong}) is the set of lines that intersect $n$
linear subspaces of codimension two in $\mathbb{P}^{n+1}$ (the union of these
subspaces constitutes the  focal variety). These subspaces can be
described explicitly: let $\lambda^{k}$ be the eigenvalues of $a$ with the
corresponding left eigenvectors $\xi^{k}$, that is,
$a^j_i\xi_j^k=\lambda^{k}\xi_i^k$.  Then the $k$-th focal subspace is defined
by two linear equations, $y^0=\lambda^k, \ \xi^k_jy^j=0$.

\medskip

\noindent{\bf Example 2.}
The simplest WDVV equation \cite{Dub2b},
$f_{ttt} = f_{xxt}^2 - f_{xxx}f_{xtt}$, can be reduced to a 3-component
conservative form,
\begin{equation}
  \label{as}
  u^1_t=u^2_x,\quad u^2_t=u^3_x,\quad u^3_t=((u^2)^2-u^1u^3)_x,
\end{equation}
by setting $u^1=f_{xxx}$, $u^2=f_{xxt}$, $u^3=f_{xtt}$. System (\ref{as}) possesses a
Hamiltonian formulation  $u_t=P{\delta H}/{\delta u}$ \cite{FN},
%\begin{equation*} \begin{pmatrix} u^1 \\ u^2 \\ u^3 \end{pmatrix}_t =P \begin{pmatrix} \fd{H}{u^1}\\ \fd{H}{u^2} \\ \fd{H}{u^3} \end{pmatrix}, \end{equation*}
with the homogeneous third-order Hamiltonian operator
  $$
  P= \ddx{}\left(
    \begin{array}{ccc}
      0 & 0 & \displaystyle\ddx{} \\
      0 & \displaystyle\ddx{} & -\displaystyle\ddx{}u^1 \\
      \displaystyle\ddx{} & -u^1\displaystyle\ddx{} &
                                                      \displaystyle\ddx{}u^2 + u^2\ddx{} + u^1 \ddx{} u^1
    \end{array}\right)
  \ddx{},
    $$
    and the nonlocal Hamiltonian
    \begin{displaymath}
      { H}=-\int\left(  \frac{1}{2}u^1(b^2)^2 + b^2b^3)\right)dx.
    \end{displaymath}
    Note that system (\ref{as}) possesses a compatible first-order Hamiltonian
    formulation, as well as a Lax pair, which elucidate its integrability
    \cite{FN}.

    The associated congruence (\ref{cong}) was thoroughly investigated in
    \cite{AF2}. It consists of trisecant lines of the focal variety which, in
    this case, is a generic projection of the Veronese surface
    $V^2\subset \mathbb{P}^5$ into $\mathbb{P}^4$. Various generalisations of
    this example can be found in \cite{KN1, KN2, pv, kkvv2}.

    \medskip

    \noindent{\bf Example 3.} The following $4$-component conservative system
    was obtained in \cite{AF3} in the classification of non-diagonalisable
    linearly degenerate systems of Temple's class whose characteristic speeds
    are harmonic (have cross-ratio equal to $-1$):
    \begin{equation}
      \label{AF1}
      \begin{array}{l}
        u^1_t=u^3_x, \\
        \ \\
        u^2_t=u^4_x,\\
        \ \\
        u^3_t=\left(\frac{u^1u^2u^4+u^3((u^3)^2+(u^4)^2-(u^2)^2-1)}{u^1u^3+u^2u^4}\right)_x,\\
        \ \\
        u^4_t=\left(\frac{u^1u^2u^3+u^4((u^3)^2+(u^4)^2-(u^1)^2-1)}{u^1u^3+u^2u^4}\right)_x.
      \end{array}
    \end{equation}
    System (\ref{AF1}) possesses a Hamiltonian representation
    $u_t=P{\delta H}/{\delta u}$ where the third-order Hamiltonian operator $P$
    is generated by the Monge metric
    \begin{displaymath}
      g_{ij}=
      \begin{pmatrix}
        (u^2)^{2} +(u^3)^{2} +1& -u^1 u^2 +u^3 u^4& -u^1 u^3 +u^2 u^4&
        -2 u^2 u^3\\
        -u^1 u^2 +u^3 u^4&(u^1)^{2} +(u^4)^{2} +1& -2 u^1 u^4&u^1 u^3
        -u^2 u^4\\
        -u^1 u^3 +u^2 u^4& -2 u^1 u^4&(u^1)^{2} +(u^4)^{2}&u^1 u^2
        -u^3 u^4\\
        -2 u^2 u^3&u^1 u^3 -u^2 u^4&u^1 u^2 -u^3 u^4&(u^2)^{2} +(u^3)^{2}
      \end{pmatrix}.
    \end{displaymath}
    Due to $\det g=(u^1u^3+u^2u^4)^2$, its singular variety consists of a
    double quadric and a double plane an infinity. The corresponding nonlocal
    Hamiltonian $H$ is given by
\begin{displaymath}
H=-\frac{1}{2}\int (b^1 b^3+ b^2b^4+ x(b^1 u^3 - u^1 b^3 +
 b^2 u^4 - u^2 b^4))dx,
\end{displaymath}
note the explicit dependence on $x$.  Integrability of system (\ref{AF1}) can be demonstrated as follows. Introducing
the $2\times 3$ matrix
$$
Z=\left(
  \begin{array}{ccc}
    u^1&u^2&1\\
    u^3&u^4&0
  \end{array}
\right),
$$
one can represent (\ref{AF1}) in matrix form (compare with Sect. 4 in
\cite{isopar}),
\begin{equation}
  Z_t=(aZZ^TZ+bZ)_x,
  \label{WW}
\end{equation}
where
$ a=\frac{1}{u^1u^3+u^2u^4}, \ b=-\frac{(u^1)^2+(u^2)^2+1}{u^1u^3+u^2u^4}.  $
Introducing the $5\times 5$ skew-symmetric matrix
$$
S=\left(
  \begin{array}{cc}
    0&Z\\
    -Z^t&0
  \end{array}
\right),
$$
one can rewrite (\ref{WW}) as a matrix Hopf-type equation,
$$
S_t=(bS-aS^3)_x,
$$
with the Lax pair
$$
\psi_x=\lambda S\psi, ~~~ \psi_t=\lambda(bS-aS^3)\psi.
$$

Congruence (\ref{cong}) associated with system (\ref{AF1}) is related to the
Cartan isoparametric hypersurface in $ S^5$, see \cite{AF3} for further
details.  Note that its focal variety is reducible.

\medskip
\noindent{\bf Example 4.}
Let us consider a class of conservative 4-component systems of the form
\begin{equation*}
  % \label{eq:8}
  u^1_t=u^2_x,\quad u^2_t=u^3_x,\quad u^3_t=u^4_x,
  \quad u^4_t=[f(u^1, \dots, u^4)]_x.
\end{equation*}
Under the substitution $u^1=f_{xxxx}$, $u^2=f_{xxxt}$, $u^3=f_{xxtt}$,
$u^4=f_{xttt}$ they reduce to a scalar fourth-order PDE for $f(x, t)$. One can
show that modulo equivalence transformations there exist only two types of such
systems possessing third-order Hamiltonian structures:

\noindent{\bf Case 1.} $f=(u^2)^2-u^1u^3$. The corresponding system
possesses a Hamiltonian formulation $u_t=P{\delta H}/{\delta u}$ with the
third-order Hamiltonian operator 
\begin{equation*}
  P =\ddx{}
  \begin{pmatrix}
    0 & 0 & 0 & \ddx{} \\
    0 & 0 & \ddx{} & 0 \\
    0 & \ddx{} & 0 & -\ddx{}u^1 \\
    \ddx{} & 0 & - u^1\ddx{} &
    \ddx{}u^2+u^2\ddx{}
  \end{pmatrix}
  \ddx{},
\end{equation*}
and the nonlocal Hamiltonian
\begin{displaymath}
  H =-\frac{1}{2}\int (u^1 (b^2)^2
  + 2b^2 b^4 +(b^3)^2)dx.
\end{displaymath}

\noindent{\bf Case 2.} $f=(u^3)^2-u^2u^4+ u^1$. The corresponding system
possesses a Hamiltonian formulation $u_t=P{\delta H}/{\delta u}$ with the
third-order Hamiltonian operator 
\begin{equation*}
  P =\ddx{}
  \begin{pmatrix}
    \ddx{} & 0 & 0 & 0 \\
    0 & 0 & 0 & \ddx{} \\
    0 & 0 & \ddx{} & -\ddx{} u^2 \\
    0 & \ddx{} & - u^2\ddx{} &
    \ddx{}g+g\ddx{}
  \end{pmatrix}
  \ddx{},
\end{equation*}
where $g=u^3+\frac{1}{2}(u^2)^2$,
and the nonlocal Hamiltonian
\begin{equation*}
  H =
  \int (b^2 b^3 u^3 -b^1 b^2  - b^3 b^4) dx.
\end{equation*}
We point out that systems from cases 1, 2 are likely to be non-integrable.

\subsection{Projective invariance}
\label{sec:proj}

The class of conservative systems (\ref{eq:61}) is invariant under reciprocal
transformations of the form
\begin{equation}
  \label{recip}
  \begin{array}{c}
    d\tilde x=(a_iu^i+a)dx+(a_iV^i+b)dt, \\
    d\tilde t=(b_iu^i+c)dx+(b_iV^i+d)dt,
  \end{array}
\end{equation}
which can be viewed as nonlocal changes of the independent variables $x, t$;
here $a_i, b_i, a, b, c, d$ are arbitrary constants. It was shown in \cite{AF1,
  AF2} that, along with affine transformations of the dependent variables
$u^i$, reciprocal transformations generate the group $SL(n+2)$ which acts by
projective transformations on the associated congruence (\ref{cong}). It is
remarkable that these transformations preserve the Hamiltonian property.

\begin{theorem}\label{invar} The class of conservative systems (\ref{eq:61})
  possessing third-order Hamiltonian formulation (\ref{casimir}) is invariant
  under reciprocal transformations (\ref{recip}).
\end{theorem}

We prove this result in Section \ref{sec:invar}. Note that, in contrast to
third-order operators (\ref{casimir}), first-order Hamiltonian structures of
Dubrovin-Novikov type are not reciprocally invariant, and generally become
nonlocal \cite{fp}.

\medskip

\noindent{\bf Remark.} In \cite{fpv, fpv1} we have classified  third-order Hamiltonian operators/systems modulo the restricted group of reciprocal transformations that change the independent variable $x$ only, 
\begin{equation*}
 % \label{recip}
  \begin{array}{c}
    d\tilde x=(a_iu^i+a)dx+(a_iV^i+b)dt, ~~~~ 
    d\tilde t=dt.
  \end{array}
\end{equation*}
In the 3-component case this resulted in the 5 canonical forms. Modulo extended transformations (\ref{recip}), all of them are equivalent to that of Example 2 from Section \ref{sec:ex}.  

\medskip
Ultimately, the classification of Hamiltonian systems of conservation laws
(\ref{eq:61}) up to reciprocal transformations (\ref{recip}) reduces to projective
classification of the associated congruences (\ref{cong}).

\subsection{Classification results}
\label{sec:class}

Here we summarise the classification results of Hamiltonian systems of
conservation laws with $n=2$ and $ 3$ components. The classification is
performed modulo reciprocal/projective transformations as discussed in Section
\ref{sec:proj}. We always assume that system (\ref{eq:61}) is strictly
hyperbolic, and that the metric $g$ defining Hamiltonian operator
(\ref{casimir}) is non-degenerate.

The existing classification of linear congruences in $\mathbb{P}^3$ and
$\mathbb{P}^4$ readily leads to the classification of $2$- and $3$-component
Hamiltonian systems of conservation laws. Thus, every linear congruence in
$\mathbb{P}^3$ consists of bisecants of two skew lines in $\mathbb{P}^3$. This
leads to

\begin{theorem} \label{n=2} For $n=2$, every Hamiltonian system of conservation
  laws is linearisable (that is, equivalent to  2-component case of Example 1
  from Section \ref{sec:ex}).
\end{theorem}

Linear congruences in $\mathbb{P}^4$ were classified by Castelnuovo in
\cite{Cas}: they can be obtained as trisecant lines of suitable projections of
the Veronese surface from $\mathbb{P}^5$ into $\mathbb{P}^4$.  Thus, all
generic linear congruences are projectively equivalent (non-generic projections
correspond to systems with degenerate Hamiltonian operators).

\begin{theorem} \label{n=3} For $n=3$, every Hamiltonian system of conservation
  laws is either linearisable (that is, equivalent to  3-component case of
  Example 1 from Section \ref{sec:ex}), or equivalent to the system of WDVV
  equations (Example 2 from Section \ref{sec:ex}).
\end{theorem}

Theorems \ref{n=2}, \ref{n=3} are proved in Section \ref{sec:class2}. It follows
that all 3-component systems of conservation laws with third-order Hamiltonian
structures are automatically integrable.

The case $n=4$ is far more complicated, primarily, due to the fact that
there exists no classification of linear congruences in $\mathbb{P}^5$. Only
partial results are currently available. In particular, 4-component Hamiltonian
systems (\ref{eq:61}) associated with third-order Hamiltonian operators are not integrable in general.

\subsection{Algebraic reformulation of the problem}
\label{sec:alg}

Linear congruences in $\mathbb{P}^{n+1}$ are defined by $n$ linear relations in
the Pl\"ucker coordinates. Setting $\mathbb{P}^{n+1}=\mathbb{P}(W)$ where $W$
is a vector space of dimension $n+2$, these linear relations correspond to the
choice of an $n$-dimensional subspace $A\subset \Lambda^2(W)$. Let
$A^1, \dots, A^n$ denote a basis of $A$. The condition that the corresponding
system (\ref{eq:61}) is Hamiltonian, is equivalent to the existence of a
non-degenerate symmetric matrix $\phi_{\beta \gamma}$, the same as in
(\ref{tri}), such that
$$
\phi_{\beta \gamma}A^{\beta}\wedge A^{\gamma}=0,
$$
see Section \ref{sec:alg1}. The existence of such relation does not depend on
the choice of  basis, and imposes strong constraints on $A$. Despite its
apparent simplicity, the classification of normal forms of such subspaces is an
open problem (starting with $n=4$).

\subsection{Symbolic computations}

Symbolic computations were performed by CDE \cite{CDE}, a Reduce \cite{reduce}
package for integrability of PDEs. CDE (by one of us, RFV) can compute:
Fr\'echet derivatives, formal adjoints, symmetries and conservation laws,
Hamiltonian operators, and their brackets. Examples are available from \cite{CDE}, and
a User's manual is included in the official Reduce manual; a book with numerous
detailed computations is to appear soon \cite{kvv-book}.

\section{Proofs}

\subsection{Conditions for a system to be Hamiltonian: proof of Theorem
  \ref{ThV}}
\label{sec:ThV}

% In this section we prove Theorem \ref{ThV}, \medskip

In this section we derive the necessary and sufficient conditions for system
(\ref{eq:61}) to possesses Hamiltonian structure (\ref{casimir}).

\medskip

\noindent {\bf Theorem \ref{ThV}.} {\it The necessary and sufficient conditions for a conservative system  (\ref{eq:61}) to possess third-order Hamiltonian operator (\ref{casimir}) are the following:
  \begin{subequations}
    \begin{align}
 \nonumber     & g_{im}V^{m}_{j}=g_{jm}V^m_{i},\\
\nonumber      & c_{mkl}V^m_{i}+c_{mik}V^m_{l}+c_{mli}V^m_{k}=0,\\
\nonumber      &V^k_{ij}=g^{ks}c_{smj}V^m_{i}+g^{ks}c_{smi}V^{m}_{j},
           \end{align}
    \end{subequations}
    here low indices of $V^m$ denote partial derivatives,
    $V^m_i=\partial V^m/\partial u^i$, etc.}\bigskip

  \begin{proof}
    The proof is based on the Kersten--Krasil'shchik--Verbovetsky approach to
    Hamiltonian operators \cite{KeKrVe-JGP-2004} which can be summarised
    as follows.  Consider an evolutionary system of the form
    \begin{equation}
      \label{evol} 
      F^i = u^i_t -f^i(t,x,u,u_{x},u_{xx},\ldots) = 0,
    \end{equation}
    with the formal linearization (Fr\'echet derivative) $\ell_F$. Let $P$ be a
    Hamiltonian operator, that is, a skew-adjoint operator with zero Schouten
    bracket, $[P, P]=0$. If system (\ref{evol}) possesses $P$ as a Hamiltonian
    structure, then $P$ maps variational derivatives of conserved densities of
    (\ref{evol}) into generalized (higher) symmetries, that is,
    \begin{equation} \label{id} \ell_F\circ P = P^*\circ\ell^*_F.
    \end{equation}
    Let us introduce the adjoint system (\emph{cotangent covering}) of system
    (\ref{evol}),
    \begin{equation}
      \left\{ 
        \begin{array}{l}
          F=0, \\ 
          \ell _{F}^{\ast }({p})=0,
        \end{array}%
      \right.  \label{cot}
    \end{equation}%
    where ${p}$ is an auxiliary (vector) variable.  Then (\ref{id}) is
    equivalent to
    \begin{equation} \label{preHam} \ell _{F}(P({p}))=0,
    \end{equation}
    which must hold identically modulo (\ref{cot}).  Note that the idea of
    representing Hamiltonian operators $P$ by linear differential expressions
    of type $P({p})$ was used in \cite{getz} to compute Hamiltonian cohomology.  The advantage of the above formulation is that finding
    Hamiltonian operators amounts to solving a problem which is computationally
    the same as finding generalized symmetries.
    
    To apply this technique to system \eqref{eq:61} we introduce a potential
    substitution $u^{i}=b_{x}^{i}$, for reasons that will become clear soon,
    obtaining a \textit{non-quasilinear} system
    \begin{equation}
      b_{t}^{i}=V^{i}(\mathbf{b}_{x}) .  \label{eq:24}
    \end{equation}
    This substitution turns Hamiltonian operator
    \eqref{casimir} into a first-order operator,
    \begin{equation} \label{eq:500} P^{ij} = -(g^{ij}(\mathbf{b}_{x})\partial
      _{x}+c_{k}^{ij}(\mathbf{b}%
      _{x})b_{xx}^{k}),
    \end{equation}
    and the corresponding Hamiltonian can be calculated explicitly (see
    Section~\ref{sec:Ham}).  Note that the above Hamiltonian operator is
    \textbf{not} of Dubrovin--Novikov type as its coefficients $g^{is}(%
    \mathbf{b}_{x})$ and $c_{k}^{is}(\mathbf{b}_{x})$ loose their geometric
    interpretation: their transformation rule is no longer tensorial.  The
    linearisation operator of system~\eqref{eq:24} is
    \begin{equation*} \label{eq:25} \ell_F(\boldsymbol{\varphi}) = D_t\varphi^i
      - \frac{\partial V^i}{\partial b^j_x}D_x\varphi^j,
    \end{equation*}
    with the adjoint
    \begin{equation*} \label{eq:27} \ell_F^*(\boldsymbol{\psi}) = -D_t\psi_k +
      D_x\left(\frac{\partial V^i}{%
          \partial b^k_x}\psi_i\right).
    \end{equation*}
    The adjoint system is
    \begin{align*}
      &b_{t}^{i}=V^{i}(\mathbf{b}_{x}), \\
      &p_{k,t} = \frac{\partial^2 V^i}{\partial b_x^k \partial b_x^h}b^h_{xx}p_i + 
        \frac{\partial V^i}{\partial b^k_x}p_{i,x}.
    \end{align*}
    Setting $P(\mathbf{p}) = -g^{ij}p_{j,x} - c^{ij}_kb^k_{xx}p_j$,
    condition (\ref{preHam}) takes the form
    \begin{multline*} \label{eq:29} \ell_F(P(\mathbf{p})) = - \frac{\partial
        g^{ij}}{\partial b^k_x} b^k_{xt}p_{j,x} - g^{ij}p_{j,xt} -
      \frac{\partial c^{ij}_k}{\partial b^h_x}%
      b^h_{xt}b^k_{xx}p_j - c^{ij}_kb^k_{xxt}p_j - c^{ij}_kb^k_{xx}p_{j,t} \\
      +\frac{\partial V^i}{\partial b^j_x} \left(\frac{\partial
          g^{jk}}{\partial b^h_{x}}b^h_{xx}p_{k,x} + g^{jk}p_{k,xx} +
        \frac{\partial c^{jh}_k}{\partial b^l_x}b^l_{xx}b^k_{xx}p_h
        +c^{jh}_kb^k_{xxx}p_h + c^{jh}_kb^k_{xx}p_{h,x} \right).
    \end{multline*}
    Using differential consequences of the adjoint system,
    \begin{gather*}
      b^i_{tx} = V^i_x,\qquad b^i_{txx} = V^i_{xx},\qquad p_{k,tx} =
      D_{xx}\frac{\partial V^i}{\partial b^k_x} p_i + 2D_x\frac{%
        \partial V^i}{\partial b^k_x} p_{i,x} + \frac{\partial V^i}{\partial
        b^k_x} p_{i,xx},
    \end{gather*}
    we obtain
    % \begin{multline*}
    %   % \label{eq:31}
    %   \ell_F(A(\mathbf{p})) =- \frac{\partial g^{ij}}{\partial b^k_x} V^k_x
    %   p_{j,x} - g^{ij}\left( D_{xx}\frac{\partial V^h}{\partial b^j_x} p_h +
    %     2D_x%
    %     \frac{\partial V^h}{\partial b^j_x} p_{h,x} + \frac{\partial
    %       V^h}{\partial
    %       b^j_x} p_{h,xx} \right) \\
    %   - \frac{\partial c^{ij}_k}{\partial b^h_x}V^h_xb^k_{xx}p_j -
    %   c^{ij}_kV^k_{xx}p_j - c^{ij}_kb^k_{xx} \left(\frac{\partial^2
    %       V^l}{\partial b_x^j \partial b_x^h}b^h_{xx}p_l + \frac{\partial
    %       V^l}{\partial b^j_x}%
    %     p_{l,x}\right) \\
    %   +\frac{\partial V^i}{\partial b^j_x} \left(\frac{\partial
    %       g^{jk}}{\partial b^h_{x}}b^h_{xx}p_{k,x} + g^{jk}p_{k,xx} +
    %     \frac{\partial c^{jh}_k}{\partial b^l_x}b^l_{xx}b^k_{xx}p_h
    %     +c^{jh}_kb^k_{xxx}p_h + c^{jh}_kb^k_{xx}p_{h,x} \right).
    % \end{multline*}
    % Collecting similar terms we have
    \begin{align*} \ell_F&(P(\mathbf{p})) =\left(-
        g^{ij}\frac{\partial V^h}{\partial b^j_x}
       + \frac{\partial V^i}{\partial b^j_x} g^{jh}\right)p_{h,xx} \\
       &+\left(- \frac{\partial g^{ih}}{\partial b^k_x} V^k_x -
         g^{ij}2D_x\frac{\partial V^h}{\partial b^j_x}
         - c^{ij}_kb^k_{xx}\frac{\partial V^h}{\partial b^j_x}\right.
         \\
         &\hphantom{+}\left.+\frac{\partial V^i}{\partial b^j_x}
         \frac{\partial g^{jh}}{\partial b^k_{x}}b^k_{xx}
         +\frac{\partial V^i}{\partial b^j_x}c^{jh}_kb^k_{xx} \right)p_{h,x} \\
      \begin{split}
        &+\Biggr(- g^{ij}D_{xx}\frac{\partial V^h}{\partial b^j_x} -
        \frac{\partial c^{ih}_k}{\partial b^j_x}V^j_xb^k_{xx} -
        c^{ih}_kV^k_{xx} - c^{ij}_kb^k_{xx}\frac{\partial^2 V^h}{\partial
          b_x^j \partial b_x^l} b^l_{xx}
        \\
        & +\frac{\partial V^i}{\partial b^j_x} \left( \frac{\partial
            c^{jh}_k}{\partial b^l_x}b^l_{xx}b^k_{xx} +c^{jh}_kb^k_{xxx}
        \right)\Biggl)%
        p_h.
      \end{split}%
    \end{align*}
    The above expression is linear in $p_h$, $p_{h,x}$, $p_{h,xx}$, and the
    coefficients are polynomials in $b^i_{xx}$, $b^i_{xxx}$. So, the expression
    vanishes if and only if
    \begin{subequations}\label{eq:46}
      \begin{align}
        \label{eq:50} &- g^{ij}\frac{\partial V^h}{\partial b^j_x} +
                        \frac{\partial V^i}{\partial b^j_x} g^{jh} = 0, \\
        \label{eq:5} & -g^{ik}\frac{\partial^2 V^h}{\partial
                       b^k_x\partial b^l_x} -c^{ih}_k\frac{%
                       \partial V^k}{\partial b^l_x} +\frac{\partial V^i}{\partial
                       b^k_x}c^{kh}_l =
                       0, \\
        \label{eq:3} & -
                       \frac{\partial g^{ih}}{\partial b^k_x} \frac{\partial V^k}{\partial
                       b^l_x%
                       } - g^{ij}2\frac{\partial^2 V^h}{\partial b^j_x\partial b^l_x} -
                       c^{ij}_l%
                       \frac{\partial V^h}{\partial b^j_x} + \frac{\partial V^i}{\partial b^j_x}
                       \frac{\partial g^{jh}}{\partial b^l_{x}} +\frac{\partial V^i}{\partial
                       b^j_x}%
                       c^{jh}_l=0, \\       \begin{split} 
                         \label{eq:6} & - g^{ij}\frac{\partial^3 V^h}{\partial
                           b^j_x\partial b^l_x\partial b^m_x} -%
                         \frac{1}{2}\left( \frac{\partial c^{ih}_m}{\partial
                             b^j_x} \frac{\partial V^j%
                           }{\partial b^l_x} + \frac{\partial
                             c^{ih}_l}{\partial b^j_x} \frac{\partial
                             V^j}{\partial b^m_x} \right) -
                         c^{ih}_k\frac{\partial^2 V^k}{\partial
                           b^l_xb^m_x} \\
                         &\hphantom{+} - \frac{1}{2}\left(
                           c^{ij}_m\frac{\partial^2 V^h} {\partial
                             b_x^j \partial b_x^l} + c^{ij}_l\frac{\partial^2
                             V^h} {\partial b_x^j
                             \partial b_x^m} \right) +\frac{1}{2}\left(
                           \frac{\partial V^i}{\partial b^j_x%
                           } \frac{\partial c^{jh}_m}{\partial b^l_x} +
                           \frac{\partial V^i}{\partial b^j_x} \frac{\partial
                             c^{jh}_l}{\partial b^m_x} \right)= 0.
                       \end{split}%
      \end{align}
    \end{subequations}
    The conditions \eqref{eq:50}, \eqref{eq:5}, \eqref{eq:3} can be simplified
    by using objects $g_{ij}, c_{ijk}$ with lower indices, leading to
    \begin{subequations}
      \label{eq:43}
      \begin{align}\label{eq:72}
        & g_{ij}\frac{\partial V^{j}}{\partial b_{x}^{h}}-\frac{\partial
          V^{j}}{\partial b_{x}^{i}}g_{jh}=0,
        \\
        \label{eq:74}
        & c_{mkl}\frac{\partial V^{m}}{\partial b_x^{i}}+c_{mik}\frac{\partial
          V^{m}}{\partial b_x^{l}}+c_{mli}\frac{\partial V^{m}}{\partial b_x^{k}}=0,\\
        \label{eq:73}
        &\frac{\partial ^{2}V^{k}}{\partial b_x^{i}\partial
          b_x^{j}}=g^{ks}c_{smj}\frac{\partial V^{m}}{\partial
          b_x^{i}}+g^{ks}c_{smi}\frac{\partial V^{m}}{\partial b_x^{j}}.
      \end{align}
    \end{subequations}
    Indeed, lowering indices in~\eqref{eq:50} leads to~\eqref{eq:72}.
    Similarly, lowering indices in~\eqref{eq:5} and using~\eqref{eq:72} leads
    to~\eqref{eq:73}. Using \eqref{eq:5} to eliminate second-order derivatives
    in~\eqref{eq:3} we get
    \begin{equation*} \label{eq:35} - \frac{\partial g^{ih}}{\partial b^k_x}
      \frac{\partial V^k}{\partial b^l_x} + 2\left(c^{ih}_k\frac{\partial
          V^k}{\partial b^l_x} -\frac{\partial V^i}{%
          \partial b^k_x}c^{kh}_l \right) - c^{ij}_l\frac{\partial V^h}{\partial b^j_x}
      + \frac{\partial V^i}{\partial b^j_x} \frac{\partial g^{jh}}{\partial b^l_{x}%
      } +\frac{\partial V^i}{\partial b^j_x}c^{jh}_l=0.
    \end{equation*}
    Lowering indices and using \eqref{eq:72} again we obtain
    \begin{equation*} \label{eq:51} (g_{ps,k} + 2c_{spk})\frac{\partial
        V^k}{\partial b^l_x} +(-g_{sk,l} - c_{skl})\frac{\partial V^k}{\partial
        b^p_x} -c_{kpl}\frac{\partial V^k}{%
        \partial b^s_x} = 0.
    \end{equation*}
    Using \eqref{eq3} we obtain \eqref{eq:74}.  It remains to show that
    equation~\eqref{eq:6} is a differential consequence of equations
    \eqref{eq:43}. In order to prove this statement we will need
    equation~\eqref{eq7}. Let us differentiate \eqref{eq:73} with respect to
    $b^l_x$ and lower the index $k$ by the metric $g$.
    % In the derivative of the above equation we may replace the Hessian of
    % $V^m$ in the right-hand side using again~\eqref{eq:73}.
    Using \eqref{eq3}, \eqref{eqc}, \eqref{g} we obtain
    \begin{multline}\nonumber
      g_{km}\frac{\partial ^{3}V^{m}}{\partial b_x^{j}\partial b_x^{i}\partial
        b_x^{l}}%
      =g^{pq}(c_{kpi}c_{qml}+c_{kpl}c_{qmi})\frac{\partial V^{m}}{\partial
        b_x^{j}}
      % \label{eq:223}
      \\
      +g^{pq}(c_{kpj}c_{qml}+c_{kpl}c_{qmj})\frac{\partial V^{m}}{\partial
        b_x^{i}}%
      +g^{pq}(c_{kpi}c_{qmj}+c_{kpj}c_{qmi})\frac{\partial V^{m}}{\partial
        b_x^{l}}.
    \end{multline}
    One can show that equation~\eqref{eq:6} can be brought to this form.  Let
    us first bring~\eqref{eq:6} to the form
    \begin{subequations}\label{eq:107}
      \begin{align}\label{eq:77}
        \frac{\partial ^{3}V^{k}}{\partial b_x^{i}\partial b_x^{j}\partial b_x^{l}}
        =& - \frac{1}{2}g_{jm}\left( 2c_{s}^{mk}g^{sq}c_{qpi}+
           \frac{\partial c_{i}^{mk}}{\partial b_x^{p}}+c_{i}^{ms}g^{kq}c_{qps}\right)
           \frac{\partial V^{p}}{\partial b_x^{l}}
        \\\label{eq:78}
         & -\frac{1}{2}g_{jm}\left( 2c_{s}^{mk}g^{sq}c_{qpl}+
           \frac{\partial c_{l}^{mk}}{\partial b_x^{p}}+c_{l}^{ms}g^{kq}c_{qps}\right)
           \frac{\partial V^{p}}{\partial b_x^{i}}
        \\\label{eq:79}
         & - \frac{1}{2}g^{kq}g_{jm}(c_{i}^{ms}c_{qpl}+c_{l}^{ms}c_{qpi})
           \frac{\partial V^{p}}{\partial b_x^{s}}
        \\\label{eq:80}
         & + \frac{1}{2}g_{sm}\left( \frac{\partial c_{i}^{sk}}{
           \partial b_x^{l}}+\frac{\partial c_{l}^{sk}}{\partial b_x^{i}}\right)
           \frac{\partial V^{m}}{\partial b_x^{j}}.
      \end{align}
    \end{subequations}
    Observe that the term~\eqref{eq:79} can be rearranged as
    \begin{align*}
      g_{jm}(c_{i}^{ms}c_{qpl}+&c_{l}^{ms}c_{qpi})\frac{\partial V^{p}}{\partial
                                 a^{s}}
      \\
      = &  g^{sb}(c_{bji}c_{qpl}+c_{bjl}c_{qpi})\frac{\partial V^{p}}{\partial a^{s}}
      \\
      =& g^{sp}(c_{bji}c_{qpl}+c_{bjl}c_{qpi})\frac{\partial V^{b}}{\partial a^{s}}
      \\
      =& -c_{qpl}\left( c_{bsj}\frac{\partial V^{b}}{\partial a^{i}}+c_{bis}\frac{%
         \partial V^{b}}{\partial a^{j}}\right) -c_{qpi}\left( c_{bsj}\frac{\partial
         V^{b}}{\partial a^{l}}+c_{bls}\frac{\partial V^{b}}{\partial a^{j}}\right),
    \end{align*}
    where we used~\eqref{eq:72} and~\eqref{eq:74}. Lowering indices
    in~\eqref{eq:107}, using
    % , simplify terms of the type $\frac{\partial c^{mk}_l}{\partial b^p_x}$
    % as
    \begin{equation}
      \label{eq:81}
      g_{hk}g_{jm}\frac{\partial c^{mk}_l}{\partial b^p_x} =
      \frac{\partial c_{hjl}}{\partial b^p_x}
      - \frac{\partial g_{hk}}{\partial b^p_x}g_{jm}c^{mk}_l
      - g_{hk}\frac{\partial g_{jm}}{\partial b^p_x}c^{mk}_l,
    \end{equation}
    as well as \eqref{eq3}, we arrive at \eqref{eq:6}. Replacing in
    (\ref{eq:43}) $b^i_x$ by $u^i$ we obtain (\ref{V}).  To finish the proof,
    it remains to note that conditions (\ref{V}) imply the existence of a
    (nonlocal) Hamiltonian, see Theorem \ref{Hamilt} of Section \ref{sec:Ham}
    for explicit formulae.
  \end{proof}

  \subsection{Involutivity of system \eqref{V}: proof of Theorem \ref{Thinv}}
  \label{sec:inv}

  In this section we establish the involutivity of system (\ref{V}), and
  estimate the number of parameters in the general solution.
  % \begin{subequations}
  %   \begin{align}
  %     \nonumber      & g_{im}V^{m}_{j}=g_{jm}V^m_{i},\\
  %     \nonumber     & c_{mkl}V^m_{i}+c_{mik}V^m_{l}+c_{mli}V^m_{k}=0,\\
  %     \nonumber &V^k_{ij}=g^{ks}c_{smj}V^m_{i}+g^{ks}c_{smi}V^{m}_{j}.
  %   \end{align}
  % \end{subequations}

  \medskip

  \noindent {\bf Theorem \ref{Thinv}.}  {\it System \eqref{V} is in
    involution. Its general solution depends on $\leq \frac{n(n+3)}{2}$
    arbitrary constants.}

  \begin{proof}

    We need to show that differentiation of first-order conditions (\ref{e1})
    and (\ref{e3}) does not lead to new first-order relations. Then, we need to
    demonstrate consistency of second-order equations (\ref{e2}). All this is a
    straightforward tensor algebra.  Differentiating
    \[
      g_{ip}V_{j}^{p}=g_{jp}V_{i}^{p},
    \]
    we obtain
    \[
      g_{ip}V_{jk}^{p}+g_{ip,k}V_{j}^{p}=g_{jp}V_{ik}^{p}+g_{jp,k}V_{i}^{p}.
    \]
    Using (\ref{e2})
    % $V_{ij}^{p}=g^{ps}[c_{smj}V_{i}^{m}+c_{smi}V_{j}^{m}]$
    we get%
    \[
      g_{ip}g^{ps}[c_{smj}V_{k}^{m}+c_{smk}V_{j}^{m}]+g_{ip,k}V_{j}^{p}=g_{jp}%
      g^{ps}[c_{smk}V_{i}^{m}+c_{smi}V_{k}^{m}]+g_{jp,k}V_{i}^{p}.
    \]
    Thus,
    \[
      c_{imj}V_{k}^{m}+c_{imk}V_{j}^{m}+g_{ip,k}V_{j}^{p}=c_{jmk}V_{i}^{m}%
      +c_{jmi}V_{k}^{m}+g_{jp,k}V_{i}^{p},
    \]
    or, relabelling indices,
    \[
      c_{ipj}V_{k}^{p}+c_{ipk}V_{j}^{p}+g_{ip,k}V_{j}^{p}=c_{jpk}V_{i}^{p}%
      +c_{jpi}V_{k}^{p}+g_{jp,k}V_{i}^{p}.
    \]
    This can be rewritten in the form
    \[
      (c_{ipj}-c_{jpi})V_{k}^{p}+(c_{ipk}+g_{ip,k})V_{j}^{p}-(c_{jpk}+g_{jp,k}%
      )V_{i}^{p}=0.
    \]
    Taking into account (\ref{eq4})
    % \[
    %   c_{ijk}=\frac{1}{3}(g_{ik,j}-g_{ij,k}),\text{ \
    % }g_{ij,k}=-c_{ijk}-c_{jik},%
    % \]
    we obtain%
    \[
      (c_{ipj}-c_{jpi})V_{k}^{p}-c_{pik}V_{j}^{p}+c_{pjk}V_{i}^{p}=0.
    \]
    Due to (\ref{eq5}) we can rewrite this as
    % Since $c_{ijk}=-c_{ikj}$ we have
    \[
      (c_{ipj}-c_{jpi})V_{k}^{p}+c_{pki}V_{j}^{p}+c_{pjk}V_{i}^{p}=0.
    \]
    Using (\ref{e3})
    % the identity \[ c_{pij}V_{k}^{p}+c_{pki}V_{j}^{p}+c_{pjk}V_{i}^{p}=0 \]
    we obtain
    \[
      (c_{ipj}-c_{jpi})V_{k}^{p}=c_{pij}V_{k}^{p}.
    \]
    It remains to note that the equality $c_{ipj}-c_{jpi}=c_{pij}$ holds
    identically due to the cyclic condition (\ref{eq6}). Thus, differentiation
    of (\ref{e1}) does not lead to new first-order relations.
    % \[
    %   c_{jpi}+c_{pij}+c_{ijp}=0.
    % \]

    \medskip

    Similarly, differentiating (\ref{e3}) we obtain
$$
c_{mkl,j}V^m_{i}+c_{mkl}V^m_{ij}+c_{mik, j}V^m_{l}+c_{mik}V^m_{lj}+c_{mli,
  j}V^m_{k}+c_{mli}V^m_{kj}=0.
$$
The substitution of (\ref{eq7}) and (\ref{e2}) gives
\begin{subequations}
  \begin{align}
    \nonumber      & -g^{pq}c_{pmj}c_{qkl}V^m_i+c_{mkl}g^{ms}(c_{spi}V^p_j+c_{spj}V^p_i)\\
    \nonumber     & -g^{pq}c_{pmj}c_{qik}V^m_l+c_{mik}g^{ms}(c_{spl}V^p_j+c_{spj}V^p_l)\\
    \nonumber      &-g^{pq}c_{pmj}c_{qli}V^m_k+c_{mli}g^{ms}(c_{spk}V^p_j+c_{spj}V^p_k)=0.
  \end{align}
\end{subequations}
Note that all terms apart from those containing $V^p_j$ cancel, leading to
$$
g^{ms}(c_{mkl}c_{spi}+c_{mik}c_{spl}+c_{mli}c_{spk})V^p_j=0.
$$
Due to (\ref{eq7}), (\ref{eq5}) this expression can be rewritten as
$$
-(c_{kpi, l}+c_{pik, l}+c_{ikp, l})V^p_j=0,
$$
which is an identity due to (\ref{eq6}).

\medskip

The compatibility of second-order relations (\ref{e2}) can be shown as follows.
Computation of the consistency condition $V_{ij,k}^{p}=V_{kj,i}^{p}$ gives
\begin{multline*}
  g^{ps}_{,k}[c_{smj}V_{i}^{m}+c_{smi}V_{j}^{m}]
  \\
  +g^{ps}[c_{smj}V_{ik}%
  ^{m}+c_{smj,k}V_{i}^{m}+c_{smi}V_{jk}^{m}+c_{smi,k}V_{j}^{m}]
\\
=g^{ps}_{,i}[c_{smj}V_{k}^{m}+c_{smk}V_{j}^{m}]
\\
+g^{ps}[c_{smj}V_{ik}%
  ^{m}+c_{smj,i}V_{k}^{m}+c_{smk}V_{ij}^{m}+c_{smk,i}V_{j}^{m}].
\end{multline*}
Cancelling terms with $V^m_{ik}$ results in a simplified expression,
\[
  g^{ps}_{,k}[c_{smj}V_{i}^{m}+c_{smi}V_{j}^{m}]+g^{ps}[c_{smj,k}V_{i}%
  ^{m}+c_{smi}V_{jk}^{m}+c_{smi,k}V_{j}^{m}]
\]%
\[
  =g^{ps}_{,i}[c_{smj}V_{k}^{m}+c_{smk}V_{j}^{m}]+g^{ps}[c_{smj,i}V_{k}%
  ^{m}+c_{smk}V_{ij}^{m}+c_{smk,i}V_{j}^{m}].
\]
Contraction with $g_{pq}$ gives
\[
  g_{pq}g^{ps}_{,k}[c_{smj}V_{i}^{m}+c_{smi}V_{j}^{m}]+c_{qmj,k}V_{i}%
  ^{m}+c_{qmi}V_{jk}^{m}+c_{qmi,k}V_{j}^{m}%
\]%
\[
  =g_{pq}g^{ps}_{,i}[c_{smj}V_{k}^{m}+c_{smk}V_{j}^{m}]+c_{qmj,i}V_{k}%
  ^{m}+c_{qmk}V_{ij}^{m}+c_{qmk,i}V_{j}^{m}.
\]
Taking into account (\ref{e2}) along with the identity
$g_{pq}g^{ps}_{,k}=-g^{sp}g_{pq,k}$ we get
\begin{multline*}
  -g^{sp}g_{pq,k}[c_{smj}V_{i}^{m}+c_{smi}V_{j}^{m}]+c_{qmj,k}V_{i}^{m}%
  \\
  +c_{qpi}g^{ps}[c_{smj}V_{k}^{m}+c_{smk}V_{j}^{m}]+c_{qmi,k}V_{j}^{m}%
\\
=-g^{sp}g_{pq,i}[c_{smj}V_{k}^{m}+c_{smk}V_{j}^{m}]+c_{qmj,i}V_{k}^{m}
\\
  +c_{qpk}g^{ps}[c_{smj}V_{i}^{m}+c_{smi}V_{j}^{m}]+c_{qmk,i}V_{j}^{m}.
\end{multline*}
Rearrangement gives
\begin{multline*}
  \lbrack c_{qmj,k}-c_{qpk}g^{ps}c_{smj}-g^{sp}g_{pq,k}c_{smj}]V_{i}%
  ^{m}
  \\
  +[c_{qpi}g^{ps}c_{smj}+g^{sp}g_{pq,i}c_{smj}-c_{qmj,i}]V_{k}^{m}%
\\
  +[c_{qmi,k}-g^{sp}g_{pq,k}c_{smi}+c_{qpi}g^{ps}c_{smk}+g^{sp}g_{pq,i}%
  c_{smk}
  \\
  -c_{qpk}g^{ps}c_{smi}-c_{qmk,i}]V_{j}^{m}=0.
\end{multline*}
Taking into account (\ref{eq4}) we obtain%
\[
  (c_{qmj,k}+g^{sp}c_{smj}c_{pqk})V_{i}^{m}-(c_{qmj,i}+g^{sp}c_{smj}%
  c_{pqi})V_{k}^{m}%
\]%
\[
  +[(c_{qmi,k}+g^{sp}c_{smi}c_{pqk})-(c_{qmk,i}+g^{sp}c_{smk}c_{pqi})]V_{j}%
  ^{m}=0,
\]
which is an identity due to (\ref{eq7}).
% \[
%   c_{mnk,l}=-g^{pq}c_{pml}c_{qnk}.
% \]

\medskip

Thus, system (\ref{V}) is in involution.  Since equations (\ref{e2}) express
all second-order partial derivatives of $V^i$, the general solution depends on
no more than $n+n^2$ parameters (values of $V^i$ and first-order derivatives
thereof). However, relations (\ref{e1}) impose $\frac{n(n-1)}{2}$ independent
constraints on first-order derivatives of $V^i$. Thus, the general solution
depends on no more than $n+n^2-\frac{n(n-1)}{2}=\frac{n(n+3)}{2}$ arbitrary
constants: the inequality is due to the extra first-order relations (\ref{e3})
that are not so easy to control. Examples show that solution spaces to
equations (\ref{V}) for different third-order Hamiltonian operators (with the
same number of components) may have different dimensions. In particular, the
maximal possible dimension, $\frac{n(n+3)}{2}$, corresponds to
constant-coefficient operators $(g_{ij}=\text{const}, \ c_{ijk}=0)$.
\end{proof}

\subsection{Integration of system \eqref{V}: proof of Theorem \ref{ThHam}}
\label{sec:sol}

It is quite remarkable that system (\ref{V}) for the fluxes $V^i$, which is a
linear involutive system with non-constant coefficients, can be integrated in
closed form. Let us recall that the metric $g$ defining Hamiltonian operator
(\ref{casimir}) can be represented in factorised form (\ref{tri}),
$ g_{ij}=\phi _{\beta \gamma }\psi _{i}^{\beta }\psi _{j}^{\gamma }, $ where
$\phi _{\beta \gamma }$ is a non-degenerate constant symmetric matrix, and
$ \psi _{k}^{\gamma }=\psi _{km}^{\gamma }u^{m}+\omega _{k}^{\gamma }; $ here
$\psi _{km}^{\gamma }$ and $\omega _{k}^{\gamma }$ are constants such that
$\psi _{km}^{\gamma }=-\psi _{mk}^{\gamma }$. These constants satisfy a set of
quadratic relations (\ref{ab}), (\ref{zac}). Using relations (\ref{tri}) --
(\ref{zac}), one can show that in the new variables $W^{\gamma}$ defined as
\begin{equation*}
  W^{\gamma}=\psi^{\gamma}_{k}V^{k},
\end{equation*}
system (\ref{V}) takes the form
\begin{subequations}
  \begin{align}
    \nonumber      & \phi_{\beta\gamma}[\psi_{ik}^{\beta}W^{\gamma}+\psi_{k}^{\beta}W_{i}^{\gamma
                     }-\psi_{i}^{\beta}W_{k}^{\gamma}]=0,\\
    \nonumber     & \phi_{\beta\gamma}[\psi_{ij}^{\beta}W_{k}^{\gamma}+\psi_{jk}^{\beta}%
                    W_{i}^{\gamma}+\psi_{ki}^{\beta}W_{j}^{\gamma}]=0,\\
    \nonumber      &W_{ij}^{\gamma}=0,
  \end{align}
\end{subequations}
where lower indices of $W^{\gamma}$ denote partial derivatives.  The last
condition implies that $W^{\gamma}$ are linear functions,
\begin{equation}\label{W}
  W^{\gamma}=\eta_{m}^{\gamma}u^{m}+\xi^{\gamma},
\end{equation}
while the first two conditions imply that the constants $\eta_{m}^{\gamma}$ and
$\xi^{\gamma}$ satisfy a linear system
\begin{equation}\label{W1}\begin{array}{c}
                            \phi_{\beta\gamma}[\psi_{ij}^{\beta}\eta_{k}^{\gamma}+\psi_{jk}^{\beta}%
                            \eta_{i}^{\gamma}+\psi_{ki}^{\beta}\eta_{j}^{\gamma}]=0, \\
                            \ \\
                            \phi_{\beta\gamma}[\psi_{ik}^{\beta}\xi^{\gamma}+\omega_{k}^{\beta}%
                            \eta_{i}^{\gamma}-\omega_{i}^{\beta}\eta_{k}^{\gamma}]=0.
                          \end{array}
                        \end{equation}
                        Thus, finding conservative Hamiltonian systems for a
                        {\it given} third-order Hamiltonian operator
                        (\ref{casimir}) is reduced to linear algebra. Conversely, given conservative system (\ref{eq:61}), the reconstruction of the associated Hamiltonian representation from  system (\ref{V}) reduces to a linear system for the coefficients of a Monge metric. The above
                        representation implies the following result.

                        \medskip
\noindent{\bf Theorem \ref{ThHam}.}   {\it For Hamiltonian system (\ref{eq:61}) the following conditions hold: 

\begin{itemize}

\item The corresponding congruence (\ref{cong})  is linear.

\item System (\ref{eq:61}) is linearly degenerate and belongs to the Temple class.

\item The fluxes $V^i$ are rational functions of the form
$$
V^i=\frac{Q^i}{\det \psi},
$$
where $\det \psi$ is a polynomial of degree $n-1$ defining the singular surface, and $Q^i$ are  polynomials of degree $n$. 

\end{itemize}}

\begin{proof}
  Linearity of the congruence can be demonstrated as follows. 
   Substituting
  $W^{\gamma}=\eta_{m}^{\gamma}u^{m}+\xi^{\gamma}$ and
  $\psi _{k}^{\gamma }=\psi _{km}^{\gamma }u^{m}+\omega _{k}^{\gamma }$ into the formula $W^{\gamma}=\psi^{\gamma}_{k}V^{k}$  we
  obtain a linear relation in the Pl\"ucker coordinates (note the skew-symmetry
  condition $\psi _{km}^{\gamma }=-\psi _{mk}^{\gamma }$),
  \begin{equation}
    \frac{1}{2}\psi _{km}^{\gamma }(u^{m}V^k-u^kV^m)+\omega _{k}^{\gamma }V^k-\eta_{m}^{\gamma}u^{m}-\xi^{\gamma}=0.
    \label{Plucker}
  \end{equation}
  This proves the linearity.  Linear degeneracy and the Temple property of
  system (\ref{eq:61}) follows from the linearity of the corresponding
  congruence \cite{AF2}.  Finally, solving the equations
  $W^{\gamma}=\psi^{\gamma}_{k}V^{k}$ for $V^k$ implies
  $V^k=\psi^k_{\gamma}W^{\gamma}$ where $\psi_{\gamma}^k$ is the inverse matrix
  to $\psi^{\gamma}_k$. Thus, $V^k$ will have $\det \psi$ in the denominator,
  while numerators will be polynomials of degree $n$.
\end{proof}

\subsection{Projective invariance: proof of Theorem \ref{invar}}
\label{sec:invar}

In this section we show that third-order Hamiltonian formalism (\ref{casimir})
is invariant under reciprocal transformations (\ref{recip}). This is in
contrast with the case of first-order Hamiltonian structures of
Dubrovin-Novikov type, which generally become nonlocal after a reciprocal
transformation \cite{Fer95, fp}.

\medskip
\noindent{\bf Theorem \ref{invar}.} {\it The class of conservative systems
  (\ref{eq:61}) possessing third-order Hamiltonian formulation (\ref{casimir})
  is invariant under reciprocal transformations (\ref{recip}).}

\medskip

\begin{proof}

  A general reciprocal transformation (\ref{recip}) can be represented as a
  composition,
$$
(x{\rm-transformation}) \circ (x \leftrightarrow t) \circ
(x{\rm-transformation}),
$$
where $x$-transformation is a reciprocal transformation changing the variable
$x$ only, and $x \leftrightarrow t$ denotes the `inversion', that is, the
interchange of $x$ and $t$. The invariance of Hamiltonian formalism
(\ref{casimir}) under $x$-transformations was established in \cite{fpv}. Thus,
it remains to show that third-order Hamiltonian formalism (\ref{casimir}) is
invariant under the inversion. Under this transformation, the new dependent
variables and the new fluxes are defined as $\tilde u^i=V^i, \ \tilde V^i=u^i$,
respectively. Recall that system (\ref{eq:61}) possesses Hamiltonian operator
(\ref{casimir}) if the following conditions are satisfied:

\begin{enumerate}

\item Metric $g$ of Hamiltonian operator (\ref{casimir}) possesses factorised
  form (\ref{tri}),
  $g_{ij}=\phi _{\beta \gamma }\psi _{i}^{\beta }\psi _{j}^{\gamma }$, where
  $\phi _{\beta \gamma }$ is a constant symmetric matrix and
  $\psi _{k}^{\gamma }=\psi _{km}^{\gamma }u^{m}+\omega _{k}^{\gamma }$; here the constants
 $\omega _{k}^{\gamma }$ and skew-symmetric $\psi _{km}^{\gamma }$ satisfy
  relations (\ref{ab}), (\ref{zac}).

\item The functions $\psi^{\gamma}_kV^k$ are linear in $u$:
  $ \psi^{\gamma}_kV^k=\eta^{\gamma}_mu^m+\xi^{\gamma}$, where the constants
  $\eta^{\gamma}_m, \ \xi^{\gamma}$ satisfy relations (\ref{W1}).

\end{enumerate}
We claim that the `inverted' system
% with the new dependent variables and the new fluxes defined as
% $\tilde u^i=V^i, \ \tilde V^i=u^i$,
is also Hamiltonian, and the metric of the transformed Hamiltonian operator is
given by
\begin{equation}
  \tilde g_{ij}=V^m_ig_{mp}V^p_j,
  \label{tildeg}
\end{equation}
% This is a consequence of the following formulae which can be verified by
% direct calculation:
note that this transformation rule is identical to that for first-order
Hamiltonian operators of Dubrovin-Novikov type. Thus, we have to demonstrate
the following:

\begin{enumerate}

\item Metric $\tilde g$ of the transformed Hamiltonian operator possesses
  factorised form
  $\tilde g_{ij}=\tilde \phi _{\beta \gamma }\tilde \psi _{i}^{\beta }\tilde
  \psi _{j}^{\gamma }$, where $\tilde \phi _{\beta \gamma }$ is a constant
  symmetric matrix, and
  $\tilde \psi _{k}^{\gamma }=\tilde \psi _{km}^{\gamma }V^{m}+\tilde \omega
  _{k}^{\gamma }$; here  $  \tilde \omega _{k}^{\gamma }$ and skew-symmetric
  $\tilde \psi _{km}^{\gamma }$  must satisfy
  relations (\ref{ab}), (\ref{zac}).

\item The expressions $\tilde \psi^{\gamma}_ku^k$ are linear in $V$:
  $\tilde \psi^{\gamma}_ku^k=\tilde \eta^{\gamma}_mV^m+\tilde \xi^{\gamma}$,
  where the constants $\tilde \eta^{\gamma}_m, \ \tilde \xi^{\gamma}$ satisfy
  relations (\ref{W1}).

\end{enumerate}
We claim that this is indeed the case, furthermore,
$$
\tilde \phi _{\beta \gamma }=\phi _{\beta \gamma }, ~~ \tilde \psi
_{km}^{\gamma }= \psi _{km}^{\gamma }, ~~ \tilde \omega _{k}^{\gamma
}=\eta_k^{\gamma},~~ \tilde \eta _{k}^{\gamma }=\omega_k^{\gamma},~~ \tilde
\xi^{\gamma}=-\xi^{\gamma},
$$
note that the constants with tilde's satisfy the same relations (\ref{ab}),
(\ref{zac}), (\ref{W1}).
% \begin{itemize} \item Metric $\tilde g$ of the transformed Hamiltonian
%   operator possesses factorised form
%   $\tilde g_{ij}=\phi _{\beta \gamma }\tilde \psi _{i}^{\beta }\tilde \psi
%   _{j}^{\gamma }$, where $\phi _{\beta \gamma }$ is a constant symmetric
%   matrix (the same as before) and
%   $\tilde \psi _{k}^{\gamma }=\psi _{km}^{\gamma }V^{m}+\eta _{k}^{\gamma }$;
%   note that $\psi _{km}^{\gamma }, \ \eta _{k}^{\gamma }$ satisfy relations
%   (\ref{ab}), (\ref{zac}).  \item The functions
%   $\tilde W^{\gamma}=\tilde \psi^{\gamma}_ku^k$ are linear in $V$:
%   $\tilde W^{\gamma}=\omega^{\gamma}_mV^m-\xi^{\gamma}$, where the constants
%   $\eta^{\gamma}_m, \ \xi^{\gamma}$ satisfy relations
%   (\ref{W1}).\end{itemize}

To establish part 1
% $\tilde g_{ij}=\tilde \phi _{\beta \gamma }\tilde \psi _{i}^{\beta }\tilde
% \psi _{j}^{\gamma }$,
we proceed as follows. Differentiating the relation
$W^{\gamma} = \psi^{\gamma}_kV^k$ with respect to $u^m$ we obtain
$\eta^{\gamma}_m=\psi^{\gamma}_kV^k_m+\psi^{\gamma}_{km}V^k$. Solving for
$V^k_m$ gives
$V^k_m=\psi_{\gamma}^k\eta^{\gamma}_m-\psi_{\gamma}^k\psi^{\gamma}_{sm}V^s$,
where $\psi_{\gamma}^k$ is the inverse matrix to $\psi^{\gamma}_k$. Thus, using
(\ref{tildeg}),
$$
\tilde
g_{ij}=V^m_ig_{mk}V^k_j=(\psi_{\gamma}^m\eta^{\gamma}_i-\psi_{\gamma}^m\psi^{\gamma}_{ri}V^r)g_{mk}
(\psi_{\tau}^k\eta^{\tau}_j-\psi_{\tau}^k\psi^{\tau}_{sj}V^s)
$$
$$
=(\psi_{\gamma}^m\eta^{\gamma}_i-\psi_{\gamma}^m\psi^{\gamma}_{ri}V^r)\psi^{\beta}_m\phi_{\beta
  \gamma}\psi^{\gamma}_k
(\psi_{\tau}^k\eta^{\tau}_j-\psi_{\tau}^k\psi^{\tau}_{sj}V^s)
$$
$$
=(\eta^{\beta}_i-\psi^{\beta}_{ri}V^r)\phi_{\beta \gamma}
(\eta^{\gamma}_j-\psi^{\gamma}_{sj}V^s)=(\eta^{\beta}_i+\psi^{\beta}_{ir}V^r)\phi_{
  \beta \gamma} (\eta^{\gamma}_j+\psi^{\gamma}_{js}V^s)=\tilde
\psi^{\beta}_{i}\phi_{ \beta \gamma}\tilde \psi^{\gamma}_{j},
$$
which is the required formula.  Finally, for part 2, it is a simple exercise to
verify that the relation
$\tilde \psi^{\gamma}_ku^k=\tilde \eta^{\gamma}_mV^m+\tilde \xi^{\gamma}$
follows from $ \psi^{\gamma}_kV^k=\eta^{\gamma}_mu^m+\xi^{\gamma}$.
\end{proof}

\subsection{Casimirs, Momentum, Hamiltonian}
\label{sec:Ham}

Given system (\ref{eq:61}) satisfying conditions (\ref{V}), in this section we
derive explicit formulae for the corresponding Casimirs, Momentum and the
Hamiltonian. To do so we introduce the substitution $u^{i}=b_{x}^{i}$
transforming system (\ref{eq:61}) into (non-quasilinear) first-order form
(\ref{eq:24}),
$$
b^i_t=V^i({\bf b}_x).
$$
In variables $b^i$, operator (\ref{casimir}) takes first-order form
(\ref{eq:500}). Using
$g_{ij}=\phi _{\beta \gamma }\psi _{i}^{\beta }\psi _{j}^{\gamma }$ we can
rewrite it in factorised form,
$$
P^{ij}=-\phi ^{\beta \gamma }\psi _{\beta }^{i}\partial _{x}\psi _{\gamma
}^{j},
$$
recall that $\psi _{\beta}^{i}$ is the inverse matrix to $\psi ^{\beta}_{i}$.

\begin{theorem}
  \label{Hamilt}
  System (\ref{eq:24}) can be represented in Hamiltonian form,
$$
b^i_t=V^i({\bf b}_x)=P^{ij}\frac{\delta {H}}{\delta b^{j}},
$$
with the local Hamiltonian
\begin{equation}
  \begin{array}{c}
    {H}=\int h\ dx=-\int  \phi_{\beta \gamma} \Big{[} \left( \frac{1}{3}\eta
    _{p}^{\gamma }\psi _{qm}^{\beta }b_{x}^{m}+\frac{1}{2}\omega _{p}^{\beta }\eta _{q}^{\gamma }\right) b^{p}b^{q}\\
    \ \\
    +x\xi^{\gamma}\left(\frac{1}{2}
    \psi _{pq}^{\beta }b^{p}b_{x}^{q}+\omega _{q}^{\beta }b^{q}\right) \Big{]} dx,
  \end{array}
  \label{ah}
\end{equation}
note the explicit $x$-dependence.  The $n$
Casimirs are given by
\begin{equation} {C}^{\alpha }=\int c^{\alpha}dx=\int \left( \frac{1}{2}\psi
    _{mk}^{\alpha }b_{x}^{k}+\omega _{m}^{\alpha }\right)
  b^{m}dx.  \label{kazimir}
\end{equation}
The Momentum has the form
\begin{equation} {M}=\int m\ dx=-\int \left( \frac{1}{3}\phi _{\beta \gamma
    }\omega _{q}^{\beta }\psi _{pm}^{\gamma }b_{x}^{m}+\frac{1}{2}\phi _{\beta
      \gamma }\omega _{p}^{\beta }\omega _{q}^{\gamma }\right)
  b^{p}b^{q}dx.  \label{pik}
\end{equation}

\end{theorem}

\noindent {\bf Remark.} In the particular case $\xi=0$, equations (\ref{W}), (\ref{W1}), (\ref{ah}) were obtained in \cite{pv}. If $\xi \ne 0$, the corresponding
Hamiltonian density $h$ has explicit $x$-dependence. It  may be more than just a curiosity that all known integrable systems (\ref{eq:61}) with Hamiltonian structure (\ref{casimir}) admit a {\it local} compatible first-order Hamiltonian 
operator  iff $h$ has no explicit $x$-dependence. 

\begin{proof}
  Using relations (\ref{W1}), one obtains the following expression for the
  variational derivative of $H$,
  \begin{equation*}
    \frac{\delta {H}}{\delta b^{j}}=-\phi _{\beta \gamma }(\psi
    _{jp}^{\beta }b_{x}^{p}+\omega _{j}^{\beta })(\eta _{q}^{\gamma }b^{q}+\xi
    ^{\gamma }x)=-\phi _{\beta \gamma }\psi^{\beta}
    _{j}(\eta _{q}^{\gamma }b^{q}+\xi
    ^{\gamma }x).
  \end{equation*}%
  Thus,
  \begin{equation*}
    \begin{array}{c}
      b_{t}^{i}=P^{ij}\frac{\delta {H}}{\delta b^{j}}=-\phi ^{\beta \gamma }\psi _{\beta
      }^{i}\partial _{x}\psi _{\gamma }^{j}\frac{\delta {H}}{\delta b^{j}}=\phi ^{\beta \gamma }\psi _{\beta
      }^{i}\partial _{x}\psi _{\gamma }^{j}
      \phi _{\mu \nu }\psi^{\mu}
      _{j}(\eta _{q}^{\nu }b^{q}+\xi
      ^{\nu }x)\\
      \ \\
      =\phi ^{\beta \gamma }\psi _{\beta
      }^{i}\partial _{x}\phi _{\gamma \nu }(\eta _{q}^{\nu }b^{q}+\xi
      ^{\nu }x)=\phi ^{\beta \gamma }\psi _{\beta}^{i}
      \phi _{\gamma \nu }(\eta _{q}^{\nu }b^{q}_x+\xi
      ^{\nu })\\
      \ \\
      =\psi _{\nu}^{i}(\eta _{q}^{\nu }b^{q}_x+\xi^{\nu })=\psi^i_{\nu}W^{\nu}=V^i({\bf b}_x),
    \end{array}
  \end{equation*}
  as required.  Similarly, variational derivatives of the Casimirs are
  \begin{equation*}
    \frac{\delta {C}^{\alpha }}{\delta b^{j}}=\psi _{jk}^{\alpha }b^k_x+\omega^{\alpha}_j=\psi _{j}^{\alpha },
  \end{equation*}%
  so that
  \begin{equation*}
    P^{ij}\frac{\delta {C}^{\alpha }}{\delta b^{j}}=-\phi ^{\beta \gamma }\psi _{\beta }^{i}\partial _{x}\psi _{\gamma }^{j}%
    \frac{\delta {C}^{\alpha }}{\delta b^{j}}=-\phi ^{\beta \gamma }\psi
    _{\beta }^{i}\partial _{x}\psi _{\gamma }^{j}\psi _{j}^{\alpha }=-\phi ^{\beta \gamma }\psi _{\beta
    }^{i}\partial _{x}\delta _{\gamma }^{\alpha }=0.
  \end{equation*}
  Finally, using (\ref{zac}), one computes variational derivatives of the
  Momentum,
  \begin{equation*}
    \frac{\delta {M}}{\delta b^{j}}=-\phi _{\beta \gamma }\psi
    _{j}^{\beta }\omega _{m}^{\gamma }b^{m}=-\phi _{\beta \gamma }\psi
    _{j}^{\beta }\partial _{x}^{-1}\psi _{m}^{\gamma }b_{x}^{m},
  \end{equation*}%
  thus,
  \begin{equation*}
    P^{ij}\frac{\delta {M}}{\delta b^{j}}=-\phi ^{\beta \gamma }\psi _{\beta }^{i}\partial _{x}\psi _{\gamma }^{j}%
    \frac{\delta {M}}{\delta b^{j}}=b^i_x,
  \end{equation*}
  as required. Note that in the original variables $u^i$, all of the above
  densities become nonlocal.
\end{proof}

\subsection{Algebraic reformulation of conditions (\ref{ab}), (\ref{zac}),
  (\ref{W1})}
\label{sec:alg1}

In this section we demonstrate that algebraic constraints (\ref{ab}),
(\ref{zac}), (\ref{W1}) can be represented in a compact invariant form which
substantially simplifies their analysis. Let us note that lines (\ref{cong})
pass through the points $y^i=u^i, \ y^{n+1}=1, \ y^{n+2}=0$ and
$y^i=V^i, \ y^{n+1}=0, \ y^{n+2}=1$, respectively. The corresponding Pl\"ucker
coordinates, which are $2\times 2$ minors of the $2\times (n+2)$ matrix
$$
\left(
  \begin{array}{ccccc}
    u^i & \dots &u^n&1&0\\
    V^i & \dots &V^n&0&1
  \end{array}
\right),
$$
can be arranged into $(n+2)\times (n+2)$ skew-symmetric matrix,
$$
Y=\left(
  \begin{array}{cc}
    U& ~~\vline \begin{array}{cc}
                  -V^1&u^1\\
                  \vdots & \vdots \\
                  -V^n&u^n
                \end{array}\\
    \hline
    \begin{array}{ccc}
      V^1& \dots &V^n\\
      -u^1 & \dots &-u^n
    \end{array}&\vline
                 \begin{array}{cc}
                   0& ~~1\\
                   -1& ~~0
                 \end{array}
  \end{array}
\right),
$$
here $U$ is the skew-symmetric matrix with entries $u^iV^j-u^jV^i$. In this
notation, relations (\ref{Plucker}) can be represented as
$$
trYA^{\gamma}=0,
$$
where $(n+2)\times (n+2)$ skew-symmetric matrices $A^{\gamma}$ are defined as
$$
A^{\gamma}=\left(
  \begin{array}{cc}
    \frac{1}{2}\psi^{\gamma}& \vline \begin{array}{cc}
                                       \omega^{\gamma}_1&\eta^{\gamma}_1\\
                                       \vdots & \vdots \\
                                       \omega^{\gamma}_n&\eta^{\gamma}_n
                                     \end{array}\\
    \hline
    \begin{array}{ccc}
      - \omega^{\gamma}_1& \dots &- \omega^{\gamma}_n\\
      -\eta^{\gamma}_1 & \dots &-\eta^{\gamma}_n
    \end{array}&~\vline
                 \begin{array}{cc}
                   0&\xi^{\gamma}\\
                   -\xi^{\gamma}&0
                 \end{array}
  \end{array}
\right),
$$
here $\psi^{\gamma}$ is the skew-symmetric matrix with entries
$\psi^{\gamma}_{ij}$. What is remarkable, relations (\ref{ab}), (\ref{zac}),
(\ref{W1}) compactify into a single relation
$$
\phi_{\beta \gamma}A^{\beta}\wedge A^{\gamma}=0,
$$
where each $A^{\gamma}$ is interpreted as a 2-form.

\subsection{Classification results: proof of Theorems \ref{n=2}, \ref{n=3}}
\label{sec:class2}

In this Section we summarise the classification of $2$- and $3$-component
Hamiltonian systems of conservation laws based on the classification of linear
congruences in $\mathbb{P}^3$ and $\mathbb{P}^4$.  \medskip

\noindent {\bf Theorem \ref{n=2}.} {\it For $n=2$, every Hamiltonian system of
  conservation laws is linearisable (that is, equivalent to  2-component case
  of Example 1 from Section \ref{sec:ex}). }

\medskip

\begin{proof}
  Every linear congruence in $\mathbb{P}^3$ consists of bisecants of two skew
  lines. Modulo projective transformations, any such congruence can be brought
  to the form
  \begin{equation*}
    y^1=u^1y^3+u^2y^4, ~~~ y^2=u^2y^3+u^1y^4,
  \end{equation*}
  where $y^i$ are homogeneous coordinates in $\mathbb{P}^3$. In the affine
  chart $y^4=1$, the skew lines in question can be defined as
  $y^3=1, \ y^1=y^2$ and $y^3=-1, \ y^1=-y^2$, respectively. The corresponding
  system of conservation laws is clearly linear,
  \begin{equation*}
    u^1_t=u^2_x, ~~~~~ u^2_t=u^1_x,
  \end{equation*}
  which is a particular case of Example 1.
\end{proof}

\noindent {\bf Theorem \ref{n=3}.} {\it For $n=3$, every Hamiltonian system of
  conservation laws is either linearisable (that is, equivalent to 
  3-component case of Example 1 from Section \ref{sec:ex}), or equivalent to
  the system of WDVV equations (Example 2 from Section \ref{sec:ex}). }

\begin{proof}

  Linear congruences in $\mathbb{P}^4$ were classified by Castelnuovo in
  \cite{Cas}.  In our presentation we follow \cite{AF2}, and use
  $(y^1: \dots :y^5)$ for homogeneous coordinates in $\mathbb{P}^4$. Over
  $\mathbb{C}$, every linear congruence in $\mathbb{P}^4$ can be brought to one
  of the four normal forms:
  \begin{itemize}
  \item Generic case: the focal variety is a generic projection of the Veronese
    surface $V^2\subset \mathbb{P}^5$ into $\mathbb{P}^4$:
$$
y^1=u^1y^4+u^2y^5, ~~~ y^2=u^2y^4+u^3y^5, ~~~ y^3=u^3y^4+((u^2)^2-u^1u^3)y^5.
$$
The corresponding system,
\begin{equation*}
  % \label{eq:8}
  u^1_t=u^2_x,\quad u^2_t=u^3_x,\quad u^3_t=((u^2)^2-u^1u^3)_x,
\end{equation*}
does not possess Riemann invariants (Example 2).
  
\item The focal variety is reducible, and consists of a cubic scroll and a
  plane which intersects the cubic scroll along its directrix:
$$
y^1=u^1y^4+u^2y^5, ~~~ y^2=u^2y^4+u^3y^5, ~~~ y^3=u^3y^4+\frac{u^2u^3}{u^1}y^5.
$$
The corresponding system,
\begin{equation*}
  % \label{eq:8}
  u^1_t=u^2_x,\quad u^2_t=u^3_x,\quad u^3_t=\left(\frac{u^2u^3}{u^1}\right)_x,
\end{equation*}
possesses one Riemann invariant. One can show that this system does not possess
non-degenerate third-order Hamiltonian structures.
\item The focal variety is reducible, and consists of a two-dimensional quadric
  and two planes which intersect the quadric along rectilinear generators of
  different families:
$$
y^1=u^1y^4+u^2y^5, ~~~ y^2=u^2y^4+u^3y^5, ~~~
y^3=u^3y^4+\frac{(u^3)^2-1}{u^2}y^5.
$$
The corresponding system,
\begin{equation*}
  % \label{eq:8}
  u^1_t=u^2_x,\quad u^2_t=u^3_x,\quad u^3_t=\left(\frac{(u^3)^2-1}{u^2}\right)_x,
\end{equation*}
possesses two Riemann invariants. One can show that this system does not
possess non-degenerate third-order Hamiltonian structures.
\item The focal variety consists of 3 planes in general position:
$$
y^1=u^1y^4+u^2y^5, ~~~ y^2=u^2y^4+u^3y^5, ~~~ y^3=u^3y^4+u^2y^5.
$$
The corresponding system is linear:
\begin{equation*}
  % \label{eq:8}
  u^1_t=u^2_x,\quad u^2_t=u^3_x,\quad u^3_t=u^2_x,
\end{equation*}
(Example 1).
\end{itemize}
Note that the number of planar components of the focal variety equals the
number of Riemann invariants of the associated system \cite{AF2}. 
\end{proof}

\section{Concluding remarks}

The classification of $n$-component Hamiltonian systems of conservation laws
has been reduced to the following algebraic problem: for a vector space $W$ of
dimension $n+2$, classify $n$-dimensional subspaces $A\subset \Lambda^2(W)$
satisfying a relation
$$
\phi_{\beta \gamma}A^{\beta}\wedge A^{\gamma}=0,
$$
where $A^{\alpha}$ is a basis of $A$ and $\phi$ is symmetric and non-degenerate. This
gives rise to the following natural questions:

\begin{itemize}

\item Classify normal forms of such subspaces $A$, at least for $n=4$. This
  would provide explicit coordinate representation of Hamiltonian systems of
  conservation laws.

\item Classify subspaces $A$ corresponding to {\it integrable} systems of
  conservation laws (note that for $n=2, 3$ all Hamiltonian systems are
  automatically integrable).  We emphasise that for $n\geq 4$ the integrability
  is no longer the case in general. We expect that Example 3 from Section
  \ref{sec:ex} will play a key role in this classification.

\end{itemize}
We hope to return to these questions elsewhere.

\section{Acknowledgements}

We thank R. Chiriv\`i, N. Hitchin, A. King, J.S. Krasil'shchik, L. Manivel, and
A.M. Verbovetsky for clarifying discussions. We acknowledge financial support
from GNFM of the Istituto Nazionale di Alta Matematica, the Istituto Nazionale
di Fisica Nucleare by IS-CSN4 \emph{Mathematical Methods of Nonlinear Physics},
and the Dipartimento di Matematica e Fisica ``E. De Giorgi'' of the
Universit\`a del Salento.  MVP's work was partially supported
by the Russian Science Foundation (grant No. 15-11-20013).

\end{document}